
\input phyzzx.tex 
\def\kyotoFIHS{\centerline{\sl Department of Fundamental Sciences}
	\centerline{\sl Faculty of Integrated Human Studies} 
          \centerline{\sl Kyoto University, Kyoto 606-01, Japan}} 
\def\kyotoFacS{\centerline{\sl Department of Physics, Faculty of 
 Science} \centerline{\sl Kyoto University, Kyoto 606-01, Japan}} 
\def\NP{Nucl.~Phys.~}
\def\PR{Phys.~Rev.~}
\def\PRL{Phys.~Rev.~Lett.~}
\def\AP{Ann.~Phys.~}
\def\PL{Phys.~Lett.~}
\def\PROG{Prog.~Theor.~Phys.~}
\def\MP{Int.~J.~Mod.~Phys.~}

\catcode`\@=11 
\newtoks\Pubnump   \let\pubnump=\Pubnump
\def\p@bblock{\begingroup \tabskip=\hsize minus \hsize
   \baselineskip=1.5\ht\strutbox \topspace-2\baselineskip
   \halign to\hsize{\strut ##\hfil\tabskip=0pt\crcr
       \the\Pubnum\crcr\the\pubnump\crcr\the\date\crcr}\endgroup}
\def\author#1{\titlestyle{\fourteencp #1}\nobreak}
\def\abstract{\par\dimen@=\prevdepth \hrule height\z@ \prevdepth=\dimen@
   \vskip 5mm \centerline{\twelverm ABSTRACT}}
\def\title#1{\vskip 1cm \titlestyle{\seventeenbf #1} \vskip\headskip }
\catcode`\@=12 
\VOFFSET = 1.2cm
\HOFFSET = .7cm
\pubnum{KUCP-0072}
\pubnump{KUNS-1302}
\date{November 1994}
\titlepage
\title{\doublespace Complex-time approach for semi-classical quantum tunneling}
\author{Hideaki Aoyama
\foot{E-mail address: aoyama@phys.h.kyoto-u.ac.jp}}
\kyotoFIHS
\vskip 2mm
\centerline{and}
\author{Toshiyuki Harano
\foot{E-mail address: harano@gauge.scphys.kyoto-u.ac.jp}}
\kyotoFacS
\vskip 0.5cm
\midinsert\narrower\noindent
\abstract
The complex-time method for quantum tunneling is studied. In one-dimensional
quantum mechanics, we construct a reduction formula for a Green function in
the number of turning points based on the WKB approximation. This formula
yields a series, which can be interpreted as a sum over the complex-time 
paths. The weights of the paths are determined.
\endinsert

\vskip 2cm
\endpage
\FIG\potshape{A potential $V(X)$ with asymptotic region I and II and
wide intermediate region III, where WKB approximation is valid.}


\def\pri{{^\prime}}
\def\bl{_\lambda}
\def\di{\Delta_{\rm i}}
\def\df{\Delta_{\rm f}}
\def\xi{x_{\rm i}}
\def\xf{x_{\rm f}}
\normalspace

The imaginary-time method is quite useful for study of the 
tunneling phenomena in quantum field theories.
Existence of the solutions of the equation of motion, instanton and bounce, 
allows one to apply the semi-classical approximation for path-integral. 
It is known to lead to valid results, which one can
confirm in the WKB approximation in quantum mechanical models.
Its range of applicability is, however, somewhat limited:
In the case when there are degenerate vacua,
the ordinary instanton calculation gives the right level-splitting
only for the lowest-lying states.
In the case of the decay of the metastable states,
the decay formula is valid for the lowest meta-stable state, 
the false ``vacuum".
These points were investigated by many authors in recent years,
especially in relation with tunneling phenomena in 
high energy scattering, namely, the baryon and lepton number
violation. \Ref\thooft{G.~'t Hooft \journal \PR &D14 (76) 3432;
A.~Ringwald \journal \NP &B330 (89) 1;
O.~Espinosa \journal \NP &B343 (90) 310.}
As a result, interacting-instanton method\Ref\ak{
H.~Aoyama and H.~Kikuchi \journal \PL &B247 (90) 75
\journal \PR &D43 (91) 1999
\journal \MP &A7 (92) 2741.}
and valley methods\REFS\yung{A.~V.Yung \journal \NP &B191 (81) 47.}
\REFSCON\kr{V.V.~Khoze and A.~Ringwald \journal \NP &B355 (91) 351.}
\REFSCON\akvalley{H.~Aoyama and H.~Kikuchi \journal \NP &B369 (92) 219.}
\refsend
became known to be useful.

Rubakov, et.~al.\Ref\rubakov{D.~T.~Son and V.~A.~Rubakov
\journal \NP &B424 (94) 55.}
proposed use of a complex-time method as an alternative.
In such a formalism, the external lines ``live" in 
real-time, while the tunneling is carried out in 
imaginary time, thus the interplay between the
initial conditions and the tunneling configuration
is expected to be straightforward,\Ref\ag{H.~Aoyama and H.~Goldberg 
\journal \PRL &188B (87) 506.} unlike the cases when the time is purely 
imaginary.

Complex-time method is studied extensively in toy models in the one-\break
dimensional quantum mechanics.
\REFS\mc{D.~W.~McLaughlin \journal J.~Math.~Phys.~ &13 (72) 1099.}
\REFSCON\bgrr{I.Bender, D.~Gomez, H.~Rothe and K.~Rothe \journal \NP 
		&B136 (78) 259.}
\REFSCON\levit{S.~Levit, J.~W.~Negele and Z.~Paltiel \journal \PR &C22 (80) 
1979.}
\REFSCON\pat{A.~Patrascioiu \journal \PR &D24 (81) 496.}
\REFSCON\le{A.~Lapedes and E.~Mottola \journal \NP &B203 (82) 58.}
\REFSCON\weiss{U.~Weiss and W.~Haeffner \journal \PR &D27 (83) 2916}
\refsend
It was argued to overcome the many shortcomings of
the imaginary time method.\Ref\boy{
D.~Boyanovsky, R.~Willey and R.Holman \journal \NP &B376 (92) 599.} 
However, even in the simple toy models, the formalism 
is full of the ambiguities and mysteries.
It is the purpose of this letter to expose those
points and solve those mysteries, thereby establishing the
sound basis of the complex-time method, on which decent
treatment of the quantum field theory should be done.

In its most successful formulation, \Ref\carlitz{
R.~D.~Carlitz and D.~A.~Nicole \journal \AP &164 (85) 411}
one considers the analytic continuation of the time-integral in 
Fourier-transform of the the fixed-energy Green functions (resolvents).
We consider the one-dimensional quantum mechanics of a particle of unit mass
in a potential $V(x)$ assumed to be smooth enough to allow WKB approximation.
We assume that it has asymptotic regions I and II 
and also in a wide intermediate region III as in Fig.\potshape.
The retarded resolvent (the fixed-energy Green function) is defined by the 
following;
$$ G^R(\xi, \xf ; E) = \bra{\xf} {1 \over E+i\delta -H} \ket{\xi}
=-i \int_0^\infty dT e^{i(E + i\delta)T} 
\bra{\xf} e^{-iHT} \ket{\xi}. 
\eqn\resolvent$$
This is useful as energy spectra is obtained from the poles and cuts of 
the above. Complex-time is introduced by deforming the $T$-integration contour 
in \resolvent : If any saddle point(s) exist in the complex $T$-plain, 
one tries to deform the integration contour so that it goes through
the saddle point(s).

Easiest to find among the saddle points are the physical ones:
Using the path-integral expression
$$ \bra{\xf} e^{-iHT} \ket{\xi} =
  i \int_{x(0)=\xi}^{x(T)=\xf} {D}x e^{iS},  \eqn\pathi$$ 
The resolvent \resolvent\ is written as a double-integral over $T$ and $x(t)$.
One then requires saddle-point condition for both $T$ and $x(t)$.
The former leads to the energy conservation law and the latter
the equation of motion.
If the initial point $\xi$ is in the allowed region,
a solution $x(t)$ starts from $t=0$ and moves along the real axis of $t$.
When it hits a turning point, it can turn around or start
moving into the forbidden region with $t$ moving along the imaginary axis.  
Corresponding to the choice of the move at turning points,
there are infinite of such saddle points.
It was proposed that all such saddle points should be taken into account
with specific weights.

Closer look of this method reveals many unanswered questions.
The complex $T$-plane is plagued with singularities and 
infinite number of saddle points.
Among saddle points, there are the ones that cannot
be obtained from the above procedure (which we shall call 
``unphysical saddle points").
Both of the physical and unphysical saddle points are 
distributed on infinite lattice structures.
How the path is deformed to avoid the singularities
and go through only the physical saddle points is unknown.  
Instead, one simply assumes that all the physical saddle points contribute.
Even so, the weight of each saddle point is a mystery.
They are determined so that results agree with that of that of the WKB 
approximation.  It is not known how they come about from the path. 

In order to investigate the situation,
we shall construct the Green function from the 
complete orthonormal set of the WKB-eigenfunctions of the hamiltonian.
For a state with the eigenvalue $\lambda$ in the continuum spectrum,
the second-order WKB approximation yields the following;
$$\psi_\lambda (x) = { 1 \over \sqrt{2\pi p(x)} } \times \cases{
	\left(A\bl e^{i \int_x^{a_1} p(x\pri) dx\pri} 
	+ B\bl e^{-i \int_x^{a_1} p(x\pri) dx\pri}\right)_, &for $x\in$ I, \crr
	\left(C\bl e^{-i \int_{b_2}^x p(x\pri) dx\pri} 
	+ D\bl e^{i \int_{b_2}^x p(x\pri) dx\pri}\right)_, 
		&for $x \in$ II, \cr}\eqn\wavef$$
where $p(x) = \sqrt{2(\lambda-V(x))}$.
Among the coefficients $A$, $B$, $C$, and $D$, only two 
are independent. Their inter-relation can be summarized in the following 
relation;\Ref\ak{H. Aoyama and M. Kobayashi \journal \PROG &64 (80) 1045.}
$$\pmatrix{A\bl \crr B\bl} = S(\lambda) \pmatrix{C\bl \crr D\bl}_. \eqno\eq$$ 
The 2$\times$2 matrix $S(\lambda)$ is determined by the 
shape of the potential $V(x)$ in the intermediate region.
The flux conservation law allows the following general parametrization;
$$S(\lambda) = \pmatrix{ e^{i\alpha} \cosh\rho & e^{i\beta} \sinh\rho \crr
e^{-i\beta} \sinh\rho & e^{-i\alpha} \cosh\rho }, \eqn\sabrho$$
where $\alpha$, $\beta$ and $\rho$ are real functions of $\lambda$.
For a given value of $\lambda$, there are two eigenfunctions.
We can show that one orthonormal choice is the following;
$$\eqalign{
	(A_\lambda, B_\lambda, C_\lambda, D_\lambda)^{(1)} &= \sqrt{2} \left(
	{e^{i\alpha} \over \cosh \rho}, \ 0, \ 1, \ 
	-e^{i(\alpha - \beta)}\tanh\rho
	\right), \crr
	(A_\lambda, B_\lambda, C_\lambda, D_\lambda)^{(2)} &= \sqrt{2} \left(
	e^{i(\alpha + \beta)}\tanh\rho, \ 1, \ 0, \ 
	{e^{i\alpha} \over \cosh \rho}
	\right).}
\eqn\choice$$ 
The Green function is written as an integration over continuous 
$\lambda$ and sum over the discrete spectrum.
This results in the following expression
in case $\xi, \xf \in $II as in Fig.\potshape; 
$$ G(\xi, \xf; E) = - |p(\xi)p(\xf)|^{-1/2} 
	e^{- \di} \left( e^{ \df}+i {R}  e^{-\df}\right) .  \eqn\nwell$$ 
where the argument is $\lambda=E$ and
${R} \equiv i S_{21}/S_{22}$ is essentially 
the (analytically-continued) reflection amplitude for the 
incoming wave from the right asymptotic region II.
The exponents are given by 
$$\Delta_{\rm i,f} = \int_{b_2}^{x_{\rm i,f}} dx |p(x)|.  \eqn\WandD$$

Due to the existence of the intermediate region III,
the matrix $S$, which connects the regions I and II,
can be written in terms of the matrices $S_1$ that connects
regions I and III and $S_2$ for III and II. 
More specifically, we apply the linear WKB connection formula
for the latter region and obtain
$$S =  S_1 \pmatrix{
      {1 \over 2} e^{- \Delta_1}\cos W_{2}
  &   e^{- \Delta_1} \sin W_{2} \cr
     -e^{ \Delta_1} \sin W_{2}
  &   2 e^{ \Delta_1}\cos W_{2}\cr}
 \eqno\eq$$
where $W_{2}$\  and $\Delta_{1}$ \ are defined by,
$$ W_{2} = \int_{a_{2}}^{b_{2}} dx p(x), \hskip 1cm
 \Delta_1 = \int_{b_1}^{a_2} dx |p(x)|.  \eqn\WandD$$
Using the above, we arrive at the following expressions;
$$iR = {i \over 2} 
{1-\tilde{R}_2 e^{2iW_2} \over 1+\tilde{R}_2 e^{2iW_2} } = 
{i\over 2} + (-i\tilde{R}_2)e^{2iW_2} + 
(-i)(-i\tilde{R}_2)^2 e^{4iW_2} + ..., \eqn\trep$$
$$-i\tilde{R}_2=-i{1-{1 \over 2}{R}_1 e^{-2 \Delta_1}
 	\over 1+{1 \over 2}{R}_1 e^{-2 \Delta_1}}
=(-i) + (iR_1)e^{-2 \Delta_1} + {i \over 2}(iR_1)^2 
e^{-4 \Delta_1} + ...,  \eqn\frep$$ 
where ${R}_1$ is the reflection amplitude at the turning point $b_1$.

\FIG\texpand{Diagrammatic view of the expansion of \trep\ in terms of 
$\tilde{R}_2$.}
\FIG\rexpand{Diagrammatic view of the expansion of \frep\ in terms of ${R}_1$.}

We prove that the reduction formula \trep\ and \frep\
is equivalent to summation over the classical complex-time paths:
Let us first look at the expansions of the resolvents in \trep.
[The convergence of this expansion is guaranteed by $\delta$.]
The phase $W_2$ is equal to $ET + S$ of the real-time solution
between the turning points $a_2$ and $b_2$.
The factor $i/2$ is the phase and weight at the turning point, where
the path approaches from the forbidden region and turns around.
When the the path hits the turning point from the
allowed region and turns around, it picks up the phase $-i$.
Thus the series can be understood as the
sum over the paths that goes through the allowed region $(a_2, b_2)$ 
different times (see Fig.\texpand).
Similarly, $\Delta_1$ is $-E\tau + S_{\rm E}$ and 
the expansion in \frep\ can be understood as the
summation over paths for the forbidden region $(b_1, a_2)$ (Fig.\rexpand).
The reflection coefficient $R_1$ itself can be expressed
as the sum over the oscillation in $(a_1, b_1)$.
Combined, they produce the sum over the different ways the paths 
goes back and forth between $(a_1, b_2)$.
We thus prove that the sum is only over the physical paths.
At the same time, the weights and phases are completely 
determined by these expansions.  
We have also looked at the case when $\xi$ and $\xf$ belong to
different asymptotic regions.
We have proven also that it can be obtained by the
sum over the physical paths with the specified formula for the
phase and the weight.

As a summary, we have shown the Green function is cast in the form
of sum only over the physical paths.
We have derived the general rule for the weights and phases for each paths.
These coincide the rules assumed by Carlitz and Nicole\refmark\carlitz
and correctly reproduce the WKB energy-level condition and the decay formula.
As the above procedure is reductive in its nature,
our proof is valid for potentials with more turning points.
Although we have used the linear WKB connection formula in this letter,
using the quadratic formula, for the purpose of investigating the lower energy
spectrum, should be straightforward,
as our method is solely based on the reduction of the matrix $S$.
We should stress that the weight $1/2$ for the turn around in the allowed 
region is problematical for the path-integral: How it is obtained from a path
that goes through all the saddle points is unknown. 
What we have done here is to justify a hybrid method, where we apply the 
path-integral method for the elementary processes and then combine then using
the knowledge of the wavefunction.
In view of the fact that the weights are now fixed
by our analysis, the path-integral method should be more seriously
considered, as this should open ways for decent treatment of the 
tunneling phenomena in field theories.
Detail of the analysis will be published in near future.\Ref\future{
H.~Aoyama and T.~Harano, {\sl Kyoto University Preprint} KUCP-75 (1994)}

\endpage\refout\endpage\figout\endpage

\input epsf
\epsfxsize=13cm
\epsfbox{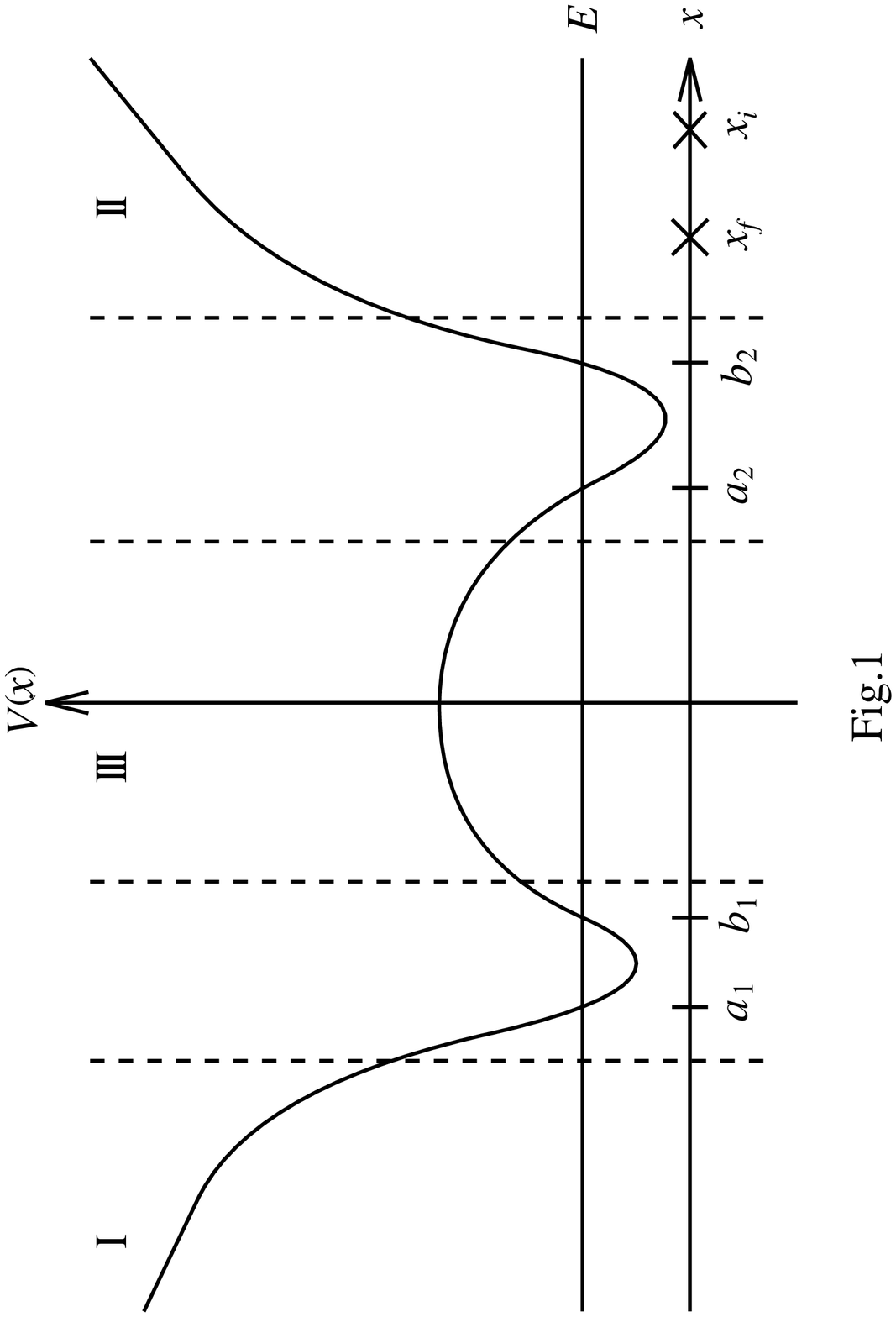}
\epsfxsize=11.8cm
\epsfbox{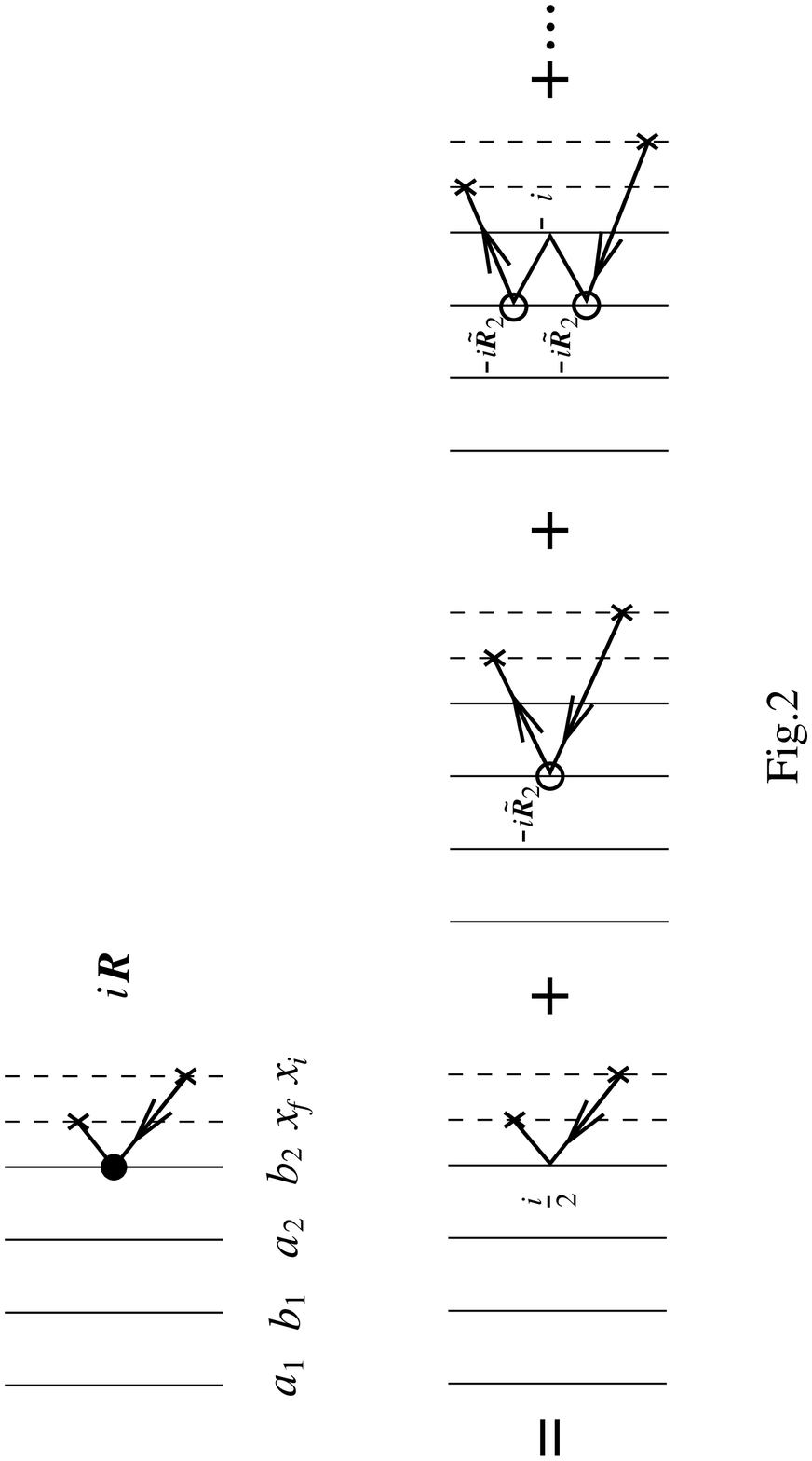}
\epsfxsize=12cm
\epsfbox{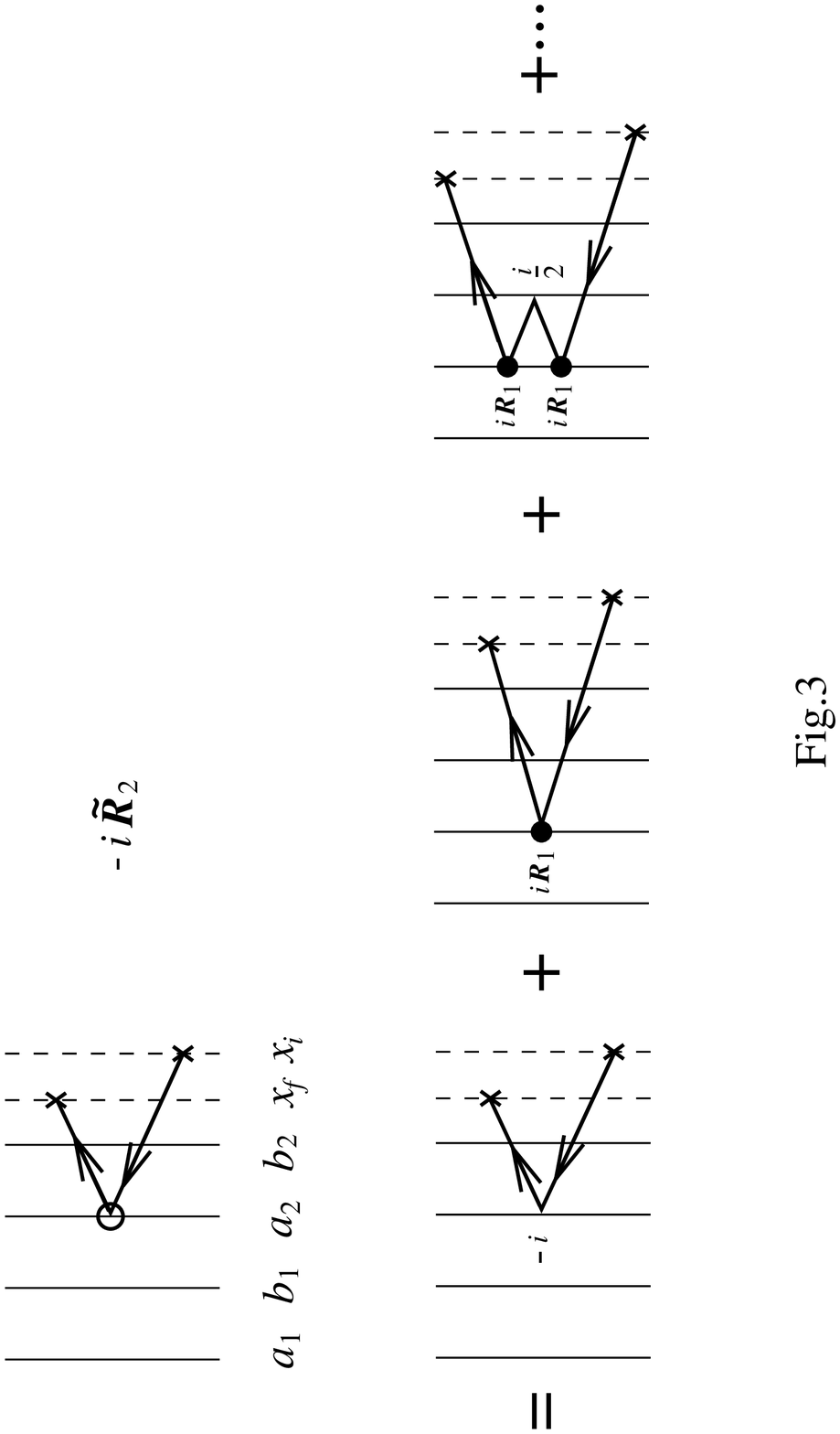}
\end